\documentclass[aps,prl,preprint,groupedaddress]{revtex4}

\usepackage[dvips]{graphicx}
\newcommand{\be}{\begin{displaymath}}
\newcommand{\bn}{\begin{equation}}

\newcommand{\en}{\end{equation}}
\newcommand{\ee}{\end{displaymath}}
\newcommand{\p}{\partial}
\newcommand{\lang}{\left\langle}
\newcommand{\rang}{\right\rangle}

\begin{document}


\title{Upper bounds on gyrokinetic instabilities}



\author{P.~Helander and G.G. Plunk}
\affiliation{Max-Planck-Institut f\"ur Plasmaphysik, 17491 Greifswald, Germany}



\date{\today}

\begin{abstract}

A family of rigorous upper bounds on the growth rate of local gyrokinetic instabilities in magnetized plasmas is derived from the evolution equation for the Helmholtz free energy. These bounds hold for both electrostatic and electromagnetic instabilities, regardless of the number of particle species, their collision frequency, and the geometry of the magnetic field. A large number of results that have earlier been derived in special cases and observed in numerical simulations are thus brought into a unifying framework. These bounds apply not only to linear instabilities but also imply an upper limit to the nonlinear growth of the free energy.

\end{abstract}


\maketitle

Most plasmas in the laboratory and in objects of astrophysical interest are strongly magnetized in the sense that the gyroradii of both electrons and ions are small in comparison with the system size. Such plasmas are subject to a wide spectrum of instabilities with wavelengths comparable to the ion or electron gyroradius, and these instabilities cause turbulence, which in turn regulates the large-scale behaviour of the plasma. 

The most complete, yet economical, mathematical framework for treating such instabilities and turbulence is provided by gyrokinetics, which has been developed since the late 1960's \cite{Taylor-1968,Rutherford} and has come to dominate large parts of theoretical plasma physics. Thousands of papers have been published on the subject \cite{Brizard-Hahm,Krommes-2012,Catto-2019}, and millions of lines of code have been written for the purpose of numerically solving gyrokinetic equations \cite{Garbet-2010}. 

On the mathematical side, the derivation of gyrokinetic equations has been discussed at great length, variational princples for these equations have been derived, conservation laws have been established, and numerical schemes respecting these laws have been devised. However, there are few {\em quantitative} mathematical results of any general validity in the field. Even something as simple as the growth rate of linear instabilities tends to be sensitive to details. A famous example is provided by the so-called ``universal'' instability of a simple plasma with a density gradient in a straight (but sheared) magnetic field. Despite its relative simplicity, such a plasma was alternately proved to be stable and unstable over the course of several decades \cite{Landreman}, and the result changes again if the geometry of the magnetic field is varied. One gets the feeling that, in gyrokinetic stability theory, the answer changes every time an assumption, however seemingly unimportant, is modified.

This state of affairs provides the motivation for the present Letter, where a family of rigorous upper bounds on the growth rates of local gyrokinetic instabilities is established. These upper bounds are valid for any collisionality, any number of particle species, and any geometry of the magnetic field. The reason these bounds are so robust is that they are derived from thermodynamic considerations, giving them a generality that is rare in the field. The results are nevertheless non-trivial, though perhaps not surprising. 

The analysis proceeds from the nonlinear gyrokinetic equation, which for each Fourier component of the ``non-adiabatic'' part, $g_{a {\bf k}}$, of the distribution function for each species $a$ reads \cite{Frieman-Chen}
	$$  \frac{\p g_{a {\bf k}}}{\p t} +
	v_{\|} \frac{\p g_{a {\bf k}}}{\p l} + i \omega_{da} g_{a {\bf k}} 
	- \frac{1}{B^2} \sum_{{\bf k}'} {\bf B} \cdot ({\bf k}' \times {\bf k}'') J_0\left( \frac{k'_\perp v_\perp}{\Omega_a} \right) 
	\left( \phi_{{\bf k}'} - v_\| A_{\| {\bf k}'} \right) g_{a {\bf k}''}
	$$
	\begin{equation} = \sum_b \left[ C_{ab}(g_{a \bf k},F_{b0}) + C_{ab}(F_{a0},g_{a \bf k}) \right] + \frac{e_a F_{a0}}{T_a} J_0\left( \frac{k_\perp v_\perp}{\Omega_a} \right)
	\left( \frac{\p}{\p t} + i \omega_{\ast a}^T \right) \left( \phi_{\bf k} - v_\| A_{\| \bf k} \right)
	\label{gk}
	\end{equation}
in the notation of  the early literature on the subject, see e.g. \cite{Antonsen,Catto-1981,Tang}. As is customary in local gyrokinetics, we consider a slender ``flux tube'' of plasma along the magnetic field and take a Fourier transform on the short scale of the perpendicular wavelength in the directions across the equilibrium field. (If the field lines trace out toroidal surfaces, this requires a ballooning transform.) The equilibrium magnetic field is written as ${\bf B} = \nabla \psi \times \nabla \alpha$, the wave vector as ${\bf k} = {\bf k}_\perp = k_\psi \nabla \psi +  k_\alpha \nabla \alpha$, ${\bf k}'' = {\bf k} - {\bf k}'$, and the equilibrium Maxwellian $F_{a0}$ is constant on surfaces of constant $\psi$. The diamagnetic frequency is denoted by $\omega_{\ast a} = (k_\alpha T_a/e_a) d \ln n_a / d \psi$, where $n_a$ represents density, $T_a$ temperature, $\eta_a = d\ln T_a / d \ln n_a$ the ratio of normalized temperature and density gradients, $e_a$ charge, and $\omega_{\ast a}^T = \omega_{\ast a} [1 + \eta_a (m_a v^2 / 2T_a - 3/2) ]$. The drift frequency is $\omega_{da} = {\bf k}_\perp \cdot {\bf v}_{da}$, where ${\bf v}_{da}$ is the magnetic drift velocity, and the linearized collision operator between species $a$ and $b$ is denoted by $C_{ab}$. The independent coordinates are the arc length $l$ along the magnetic field $\bf B$, the magnitude $v = |{\bf v}|$ of the velocity vector, and the magnetic moment $\mu_a = m_a v_\perp^2 / 2B$, where $v_\perp = | {\bf v} - {\bf v} \cdot {\bf B} {\bf B}/B^2|$. Finally, $J_0$ denotes a Bessel function and $\Omega_a = e_a B / m_a$. The electrostatic and magnetic potentials (in the Coulomb gauge) are given by the field equations
	\bn \sum_a \lambda_a \phi_{\bf k} = \sum_a e_a \int g_{a {\bf k}} J_0 d^3v, 
	\label{field1}
	\en
	\bn A_{\| {\bf k}} = \frac{\mu_0}{k_\perp^2} \sum_a e_a \int v_\| g_{a {\bf k}} J_0 d^3v, 
	\label{field2}
	\en
where $ \lambda_a = {n_a e_a^2}/{T_a}$, and for simplicity we neglect parallel magnetic-field fluctuations and any equilibrium flow. The latter is, in other words, assumed to be small enough that Coriolis and centrifugal forces can be neglected. 

Our primary aim is to derive an upper bound on the growth rate of linear instabilities as a function of plasma parameters, instability wavelength, and magnetic-field geometry. For this purpose we consider the entropy budget of the system by operating on Eq.~(\ref{gk}) with	
	$$ {\rm Re} \; \sum_{a,{\bf k}} T_a \lang \int \left( \cdots \right) \frac{g_{a \bf k}^\ast}{F_{a0}} d^3v \rang, $$
where angular brackets denote a volume average over the flux tube and an asterisk the complex conjugate. This operation annihilates many of the terms of the gyrokinetic equation and results in the relation
	\bn \sum_{\bf k} \frac{\p H}{\p t} = 2 \sum_{\bf k} (C + D), 
	\label{entropy balance}
	\en
where 
	\bn C({\bf k},t) = {\rm Re} \; \sum_{a,b} T_a \lang \int \frac{g_{a \bf k}^\ast}{ F_{a0}} 
	\left[C_{ab}(g_{a \bf k},F_{b0}) + C_{ab}(F_{a0},g_{b \bf k}) \right] d^3v \rang \le 0 
	\label{H-theorem}
	\en
is negative or vanishes by Boltzmann's H-theorem \cite{Sugama-2009}, and we have written
	$$ D({\bf k}, t) = {\rm Im} \; \sum_a  e_a \lang \int g_a \omega_{\ast a}^T 
	\left( \phi^\ast_{\bf k} - v_\| A_{\| \bf k}^\ast \right) J_0 d^3v \rang, $$
	$$ H({\bf k},t)  = \sum_a \lang  T_a \int \frac{|g_{a \bf k}|^2}{F_{a0}} d^3v - \lambda_a |\phi_{\bf k}|^2 \rang
	+ \lang \frac{| k_\perp A_{\| \bf k} |^2}{\mu_0} \rang. $$
$D({\bf k}, t)$ is related to the entropy production caused by the particle and heat fluxes associated with the perturbations $g_{a\bf k}$, and the expression for $H$ becomes intelligible when written in terms of the function $\delta F_a = g_a - (e_a J_0 \phi / T_a) F_{a0}$, which makes it clear that the quantity $H$ is, in fact, the Helmholtz free energy of the fluctuations,
	\bn H = U - \sum_a T_a S_a 
	\label{Helmholtz}
	\en
Here $ S_a = - \lang  n_a s_a \rang $ denotes the entropy perturbation of species $a$, where 
	$$ s_a = \frac{1}{n_a} \int \frac{|\delta F_{a \bf k} |^2}{F_{a0}} d^3v $$
is derived from the expansion of the Gibbs entropy $-\int F_a \ln F_a d^3v$ around a Maxwellian, and
	$$ U = \lang  \sum_a \lambda_a
	\left(1-\Gamma_0(k_\perp^2 \rho^2_a) \right) |\phi_{\bf k}|^2+ \frac{| k_\perp A_{\| \bf k} |^2}{\mu_0} \rang  $$
represents the energy of the fluctuations, where $\Gamma_n(x) = I_n(x) e^{-x}$ and $\rho^2_a = T_a / (m_a \Omega_a^2)$. The first term in $U$ is the gyrokinetic generalization of the kinetic energy of ${\bf E} \times {\bf B}$ motion, and the second term the energy associated with magnetic-field fluctuations.

In fully developed turbulence, a relation similar to $C + D = 0$ holds on a time average and has often been invoked in discussions of turbulent cascades \cite{Krommes,Brizard,Sugama,Garbet-2005,Schekochihin,Banon,SD}. Our aims are different, and instead draw inspiration from little-known work going back to Fowler \cite{Fowler-1964,Fowler-1968,Brizard-1991}. At first, we consider a single linear eigenmode and note that Eqs.~(\ref{entropy balance})-(\ref{H-theorem}) imply
	\bn \gamma \le \frac{D}{H} 
	\label{basic upper bound}
	\en
for the growth rate $\gamma(k_\psi,k_\alpha)$. Furthermore, we use the triangle and Schwarz inequalities to find an upper bound on the free-energy production $D$ as a function of the fluctuation amplitudes $\delta F_{a \bf k}$, $|\phi_{\bf k}|$ and $|A_{\| \bf k} |$,
	$$ D \le \sum_a |e_a| |S_a|^{1/2} 
	\lang \int F_{a0} \omega_{\ast a}^2 J_0^2 \left(|\phi_{\bf k}  |^2 + v_\|^2 |A_{\| \bf k}|^2 \right) d^3v \rang^{1/2}  $$
	\bn = \sum_a |e_a \omega_{\ast a}| |n_a S_a|^{1/2} 
	\lang M(\eta_a, b_a) |\phi_{\bf k}  |^2 +
	N(\eta_a, b_a) \frac{T_a |A_{\| \bf k}|^2}{m_a}  \rang^{1/2}, 
	\label{D}
	\en
where $b_a = k_\perp^2 \rho_a^2$ and 
	$$ M(\eta,b) = \left( 1 + \frac{3 \eta^2}{2} - 2\eta(1+ \eta) b + 2 \eta^2 b^2 \right) \Gamma_0(b)
	+ \eta b\left( 2 + \eta - 2 \eta b \right) \Gamma_1(b), $$
	$$ N(\eta,b) = \left( 1 + 2 \eta + \frac{7 \eta^2}{2} - 2\eta(1+ 2\eta) b + 2 \eta^2 b^2 \right) \Gamma_0(b)
	+ \eta b\left( 2 + 3 \eta - 2 \eta b \right) \Gamma_1(b). $$

The triangle and Schwarz inequalities can also be applied directly to the field equations (\ref{field1}) and (\ref{field2}), written in terms of $\delta F_a$, to infer upper bounds on the fluctuation amplitudes in terms of the entropy perturbations. In each point, we have
	\bn \sum_a \lambda_a \left(1 - \Gamma_{0a} \right) |\phi_{\bf k}  | \le \sum_a n_a |e_a| \sqrt{ \Gamma_{0a} s_a} , 
	\label{bound on phi}
	\en
	\bn \frac{k_\perp | A_{\| \bf k} |}{B} \le \sum_a \frac{\beta_a}{2 k_\perp \rho_a} \sqrt{\Gamma_{0a} s_a},
	\label{bound on A}
	\en
where we have written $\Gamma_{0a}(l) = \Gamma_0(k_\perp^2 \rho_a^2)$ and $\beta_a(l) = 2 \mu_0 n_a T_a / B^2$. From these inequalities and Eqs.~(\ref{basic upper bound})-(\ref{D}) it is possible to derive a family of rigorous upper bounds on the growth rate $\gamma$ of any instability governed by Eqs.~(\ref{gk}) - (\ref{field2}). 	

As a first example, let us consider a hydrogen plasma with Boltzmann-distributed (so-called ``adiabatic'') electrons, $g_e = 0$, which is the traditional simplest gyrokinetic model of the ion-temperature-gradient (ITG) instability. Both the curvature-driven branch and the ``slab'' branch of the instability, and any mixture thereof, are described by this model \cite{Biglari,Romanelli,Plunk}, which has been the subject of hundreds of publications. In this case, the free energy becomes
	$$ H = n T_i \lang s_{i} + \left(1 + \tau - \Gamma_{0i} \right) \left| \frac{e \phi_{\bf k}}{T_i} \right|^2 \rang $$
where $\tau=T_i/T_e$, and the bound (\ref{bound on phi}) is replaced by the more stringent condition
	$$ \left(1 +\tau - \Gamma_{0i} \right) \frac{e |\phi_{\bf k}|}{T_i} \le \sqrt{ \Gamma_{0i} s_{i}}.$$
Minimizing $H$ subject to this constraint gives
	$$ H \ge \lambda_i \left( 1 + \tau \right) \lang \left(\frac{1 + \tau}{\Gamma_{0i}} - 1 \right) \left| \phi_{\bf k} \right|^2 \rang = H_{\rm min}({\bf k}) $$
and implies $H \ge \sqrt{H_{\rm min}({\bf k}) n T_i \lang s_{i} \rang}$, which together with (\ref{D}) can be used in (\ref{basic upper bound}) to derive the inequality
	$$ \frac{\gamma}{\omega_{\ast i}} \le 
	\frac{\lang M(\eta_i, b_i) |\phi_{\bf k} |^2 \rang^{1/2}}{ \lang (1+\tau) [(1+\tau) \Gamma_{0i}^{-1} - 1]|\phi_{\bf k} |^2 \rang^{1/2}}. $$
Thanks to the peculiar property of adiabatic electrons that the particle transport vanishes identically, this bound can, in fact, be sharpened. The density gradient does not contribute to the entropy production and can therefore be removed from $D$ from the outset, causing the function $M(\eta,b)$ to be replaced by
	$$ \tilde{M}(\eta,b) = \eta^2 \left[ \left(\frac{3 }{2} - 2b + 2 b^2 \right) \Gamma_0(b)
	+ b\left(1 - 2  b \right) \Gamma_1(b) \right], $$
Since $\tilde M(\eta,b)$ and $\Gamma_0(b)$ are both monotonically decreasing functions of $b$, we thus obtain
	\bn \frac{\gamma}{\omega_{\ast i}} \le \sqrt{ \frac{ \tilde M(\eta_i, b_{\rm min})}
	{(1+\tau) \left[ (1 + \tau) \Gamma_0^{-1} (b_{\rm min}) - 1 \right]}}, 
	\label{ae}
	\en
where $b_{\rm min} = b_i(l_0)$ denotes the smallest value of $b_i(l) = k_\perp^2 \rho_i^2 \propto (k_\perp/B)^2$ anywhere along the flux tube for the pair of wave numbers $(k_\psi,k_\alpha)$ under consideration. Equation (\ref{ae}) represents a universal upper bound on all gyrokinetic instabilities (not only ITG modes but also trapped-ion modes) in a plasma with adiabatic electrons. This bound, which is plotted in Fig.~1, holds for any collisionality and any magnetic flux tube geometry, where the latter affects the result only through the variation of $k_\perp \rho_i$ along the field. For long wavelengths, $k_\perp \rho_i \ll 1$, the dependence on geometry disappears and we simply obtain	
	$$ \gamma \le |\eta_i \omega_{\ast i}| \sqrt{ \frac{3}{2\tau(1+\tau)}}. $$
Note that $\omega_{\ast i}$ is proportional to $k_\alpha$, and that the growth rate thus vanishes in the limit of long wavelength, i.e. $\gamma \rightarrow 0$ as $k_\perp \rho_i \rightarrow 0$, as invariably observed in numerical simulations. A well-known unfavorable dependence on the electron temperature is also present, which causes the growth rate to increase with increasing $T_e/T_i$ \cite{Biglari,Romanelli,Plunk,Zocco}. In the opposite limit of short wavelength, $k_\perp \rho_i \gg 1$, the bound remains finite, 
	$$ \gamma \le  \frac{| \eta_i \omega_{\ast i} |}{1+\tau} 
	\sqrt{ \frac{ 5}{8 \pi b_{\rm min}} }, $$
since $b_{\rm min}$ is positive definite and quadratic in $k_\psi$ and $k_\alpha$. Note that $\gamma(k_\psi,k_\alpha)$ approaches a finite constant in the limit $k_\alpha \rightarrow \infty$ and vanishes when $k_\psi \rightarrow \infty$. 

\begin{figure}
  \centerline{\includegraphics{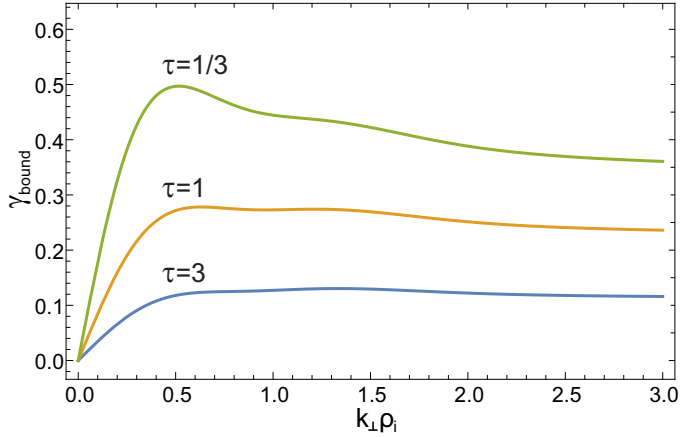}}
  \caption{Upper bound (\ref{ae}) on the growth rate (in arbitrary units) of gyrokinetic instabilities for $k_\psi = 0$ and three different values of $\tau = T_i/T_e$ in a plasma with adiabatic electrons as a function of the smallest value of $k_\perp \rho_i$ along the magnetic field.}
\label{fig:figure1}
\end{figure}

Guided by these results, we now turn to the general case of electromagnetic instabilities in a plasma with an arbitrary number of ion species. We begin by seeking lower bounds on $H$ under the constraints (\ref{bound on phi}) and (\ref{bound on A}). From the former we obtain 
	$$ H \ge \lang L |\phi_{\bf k} |^2 \rang $$
with 
	$$ L(l) = \left( \sum_a \lambda_a \right) \left( \sum_b \lambda_b (1-\Gamma_{0b}) \right) \bigg\slash \left( \sum_c \lambda_c \Gamma_{0c} \right), $$
and from the latter 
	$$ H \ge \lang \frac{|k_\perp A_{\| \bf k} |^2}{\mu_0} 
	\left[ 1 + \left( \sum_a \frac{\beta_a \Gamma_{0a}}{2 b_a} \right)^{-1} \right] \rang
	\simeq \lang \frac{|k_\perp A_{\| \bf k} |^2}{\mu_0} 
	\left( 1 + \frac{2 b_e}{\beta_e \Gamma_{0e}} \right) \rang, $$
where we have recognized that, to an excellent approximation, the sum over species is dominated by the contribution from the electrons. (The error is of order $m_e T_e / m_i T_i$.) Using these inequalities as well as $H \ge \sqrt{n_a T_a \lang s_a \rang}$ in (\ref{basic upper bound}) gives
	\bn \gamma \le \sum_a |\omega_{\ast a}| 
	\sqrt{\frac{\lang \lambda_a M(\eta_a, b_a) |\phi_{\bf k}|^2 \rang}{\lang L |\phi_{\bf k}|^2 \rang}} 
	+ |\omega_{\ast e}| 
	\sqrt{\frac{\lang N(\eta_e, b_e) |A_{\| \bf k} |^2 \rang}{\lang K |A_{\| \bf k} |^2\rang}} 
	\label{em bound}
	\en
with 
	$$ K(l) = \frac{2 b_e}{\beta_e} 
	\left( 1 + \frac{2 b_e}{\beta_e \Gamma_{0e}} \right), $$
where we again have neglected the contribution from ions to the electromagnetic term, thus committing a very small error. Since $L$ is an increasing function of $k_\perp / B$, the first term on the right of (\ref{em bound}) is maximized if $|\phi_{\bf k}(l)|^2$ is chosen to be delta function in the point $l_0$ where this quantity attains its minimum. Similarly, the second term is maximized by choosing $| A_{\| \bf k}(l) |^2 \propto \delta(l-l_1)$ where $l_1$ is the point where $K/N$ is minimized. Equation (\ref{em bound}) thus implies the upper bound
  \bn \gamma \le \sum_a |\omega_{\ast a}| 
	\sqrt{\frac{\lambda_a M(\eta_a, b_a(l_0))}{L(l_0)}} 
	+ |\omega_{\ast e}| 
	\sqrt{\frac{ N(\eta_e, b_e(l_1))}{K(l_1) }}, \label{most general bound}
	\en
which represents our most general result. Since the right-hand side is a bounded function of $k_\psi$ and $k_\alpha$, it implies an absolute upper bound on the growth rate for any wave numbers. This bound, which is conservative and can be improved by a factor of order unity at the expense of increased complexity, depends on the density and temperature gradients of all species in a non-trivial way. In the important special case of a pure hydrogen plasma and $k_\perp \rho_e \ll 1$, it reduces to 
	$$ \frac{\gamma}{|\omega_{\ast e}|} \le \sqrt{\frac{\tau(\Gamma_{0i} + \tau)}{(1 + \tau) (1 - \Gamma_{0i})} }
	\left( \sqrt{\tau M(\eta_i, b_i)} + \sqrt{1 + \frac{3 \eta_e^2}{2}} \right) 
	+ \beta_e \sqrt{\frac{ 1 + 2 \eta_a + 7 \eta_e^2/2}{2 b_e \left(\beta_e + 2 b_e \right)}}, $$
where the first term on the right is evaluated at $l=l_0$ and the second one at $l=l_1$. In the opposite limit of $k_\perp \rho_e \gg 1$, the electromagnetic term can be neglected altogether and we instead obtain
	$$ \frac{\gamma}{|\omega_{\ast e}|} \le \frac{\tau}{1 + \tau} \sqrt{\frac{1-\eta_e + 5 \eta_e^2/4}{2 \pi b_e(l_0)}}. $$
Several well-known features from gyrokinetic theory and simulations are manifest in these expressions. For instance, the contribution from magnetic fluctuations is proportional to $\beta_e$ and peaks at long wavelengths, whereas the electrostatic terms are independent of $\beta$ and increase with wave number but remain bounded as $k_\perp \rho_a \rightarrow \infty$. Heavy particle species contribute more at small wave numbers while electrons dominate at large ones, and a number of scalings with respect to the ion and electron temperatures that have earlier been derived in special cases \cite{Tang,Biglari,Romanelli,Plunk,Zocco} are also reflected in these results. 

Although the bounds (\ref{ae}) and (\ref{em bound}) have been derived for linear instabilities, they have much more general implications and we therefore now consider the nonlinear growth of free energy associated with an arbitrary initial condition, defined by distribution functions $\delta F_{a \bf k}$ at $t=0$, say. This initial condition need not correspond to a linear eigenmode and could, for instance, describe a turbulent spectrum of large-amplitude disturbances. When several Fourier modes with different wave vectors ${\bf k}$ are present, the total free energy $H_{\rm tot}$ and entropy production $D_{\rm tot}$ are equal to sums of the corresponding quantities for each wave number,	
  $$ H_{\rm tot}(t) = \sum_{{\bf k}} H({\bf k},t), $$
	$$ D_{\rm tot}(t) = \sum_{{\bf k}} D({\bf k},t), $$
where each component satisfies the bounds (\ref{ae}) or (\ref{em bound}) derived above, i.e., $ D({\bf k},t) \le \gamma_{\rm bound}({\bf k}) H({\bf k},t)$. 
According to Eq.~(\ref{entropy balance}), the nonlinear growth is thus limited by
	$$  \frac{d H_{\rm tot}}{dt} \le 2 \sum_{{\bf k}} \gamma_{\rm bound}({\bf k}) H({\bf k},t). $$
The circumstance that, according to Eqs.~(\ref{ae}) and (\ref{em bound}), the function $\gamma_{\rm bound}({\bf k})$ is itself bounded, i.e., there is a number $\gamma_{\rm max}$ such that
	$$\gamma_{\rm bound}({\bf k}) < \gamma_{\rm max} \quad \mbox{for all } {\bf k}, $$ 
implies a similar bound on the nonlinear growth of the total free energy,
		$$  \frac{d \ln H_{\rm tot}}{dt} \le 2 \gamma_{\rm max}. $$
Our bounds on linear instability growth rates thus imply a universal bound on the {\em nonlinear} growth of free energy. The latter can never exceed twice the largest linear growth rate bound, no matter how turbulent the plasma is.

Due to Boltzmann's H-theorem (\ref{H-theorem}), collisions always dissipate free energy (\ref{entropy balance}) and can only act to reduce the upper bounds that we have derived. This is somewhat curious since collisions sometimes act destabilizing in linear stability theory, but apparently such behavior is reflected in the bounds. 

Conversely, if collisions are absent, it is always possible to achieve a positive instantaneous growth rate $d\ln H_{\rm tot} / dt > 0$ of free energy by an appropriate choice of initial conditions $\delta F_{a \bf k}$ at $t=0$. This statement holds even if the system is linearly stable. (The growth will then be transient and followed by damping.) To see this, it is sufficient to note that, without collisions, the growth rate is given by $d\ln H_{\rm tot} / dt = 2 D_{\rm tot} / H_{\rm tot}$, which is a ratio of two functionals that are quadratic in the distribution functions $\delta F_{a\bf k}$ and can always be made positive for some choice of these functions. The circumstance that the free energy can grow transiently in the face of linear stability means that sub-critical turbulence is possible \cite{Landreman-Plunk-Dorland}, at least if the growth is vigorous enough, and it is therefore of importance that it cannot exceed the bounds derived above. As will be shown in a future publication, it is possible to make these 'tight' by identifying the distribution functions $\delta F_{a\bf k}$ that maximize the ratio, which then defines the largest possible rate of free-energy growth. 

In summary, a family of universal upper bounds on the linear growth rate has been found for any instability described by the gyrokinetic system of equations (\ref{gk}) - (\ref{field2}), including ion- and electron-temperature-gradient modes, the so-called ``universal'' and ``ubiquitous'' instabilities, dissipative and collisionless trapped-particle modes, kinetic and resistive ballooning modes, and micro-tearing modes. These bounds hold for plasmas consisting of any number of particle species having any collision frequency, and the magnetic geometry is also general, except for the local approximation made in the formulation of the equations themselves. A large number of results that have earlier been derived in special cases or observed in numerical simulations are thus brought into a unifying framework. Moreover, the nonlinear growth of free energy is also limited by the maximum bound on the linear growth rate. 

This work was partly supported by a grant from the Simons Foundation (560651, PH). 
		
\normalsize


\begin{references}



\bibitem{Taylor-1968} J.B. Taylor and R.J. Hastie, Plasma Phys. {\bf 10}, 479 (1968). 

\bibitem{Rutherford} P.H. Rutherford and E.A. Frieman, Phys. Fluids {\bf 11}, 569 (1968).

\bibitem{Brizard-Hahm} A. Brizard and T.S. Hahm, Rev. Mod. Phys. {\bf 79}, 421 (2007).

\bibitem{Krommes-2012} J.A. Krommes, Ann. Rev. Fluid Mech. {\bf 44}, 175-201 (2012).

\bibitem{Catto-2019} P.J. Catto, J. Plasma Phys. {\bf 85}, 925850301 (2019).

\bibitem{Garbet-2010} X. Garbet, Y. Idomura, L. Villard and T.H. Watanabe, Nucl. Fusion {\bf 50}, 043002 (2010).

\bibitem{Landreman} M. Landreman, T.M. Antonsen, Jr., and W. Dorland, Phys. Rev. Lett. {\bf 114}, 095003 (2015).

\bibitem{Frieman-Chen} E.A. Frieman and L. Chen, Phys. Fluids {\bf 25}, 502 (1982).

\bibitem{Antonsen} T.M. Antonsen and B. Lane, Phys. Fluids {\bf 23}, 1205 (1980). 

\bibitem{Catto-1981} P.J. Catto, W.M. Tang and D.E. Baldwin, Plasma Phys. {\bf 23}, 263 (1981). 

\bibitem{Tang} W.M. Tang, J.W. Connor and R.J. Hastie, Nucl. Fusion {\bf 20}, 1439 (1980). 

\bibitem{Sugama-2009} Strictly speaking, the H-theorem only holds for the exact collision operator if all particle species have equal temperatures, and we assume this to be the case for species with comparable masses. Unequal temperatures are only allowed (and, usually, expected) for particle species with widely disparate masses (such as electrons and ions), in which case the collision operator can be approximated by a form that satisfies an H-theorem. This issue is further discussed by H. Sugama, T.-H- Watanabe and M. Nunami, Phys. Plasmas {\bf 16}, 112503 (2009).

\bibitem{Krommes} J.A. Krommes and G. Hu, Phys. Fluids B {\bf 5}, 3908 (1993).

\bibitem{Brizard} A. Brizard, Phys. Plasmas {\bf 1}, 2473 (1994). 

\bibitem{Sugama} H. Sugama, M. Okamoto, W. Horton and M. Wakatane, Phys. Plasmas {\bf 3}, 2379 (1996). 

\bibitem{Garbet-2005} X. Garbet, N. Dubuit, E. Asp, Y. Sarazin, C. Bourdelle, P. Ghendrih, and G. T. Hoang,  Phys. Plasmas {\bf 12}, 082511 (2005). 

\bibitem{Schekochihin} A.A. Schekochihin, S.C. Cowley, W. Dorland, G.W. Hammett, G.G. Howes, E. Quataert and T. Tatsuno, Astrophys. J. {\bf 182}, 310 (2009). 

\bibitem{Banon} A. Banon Navarro, P. Morel, M. Albrecht-Marc, D. Carati, F. Merz, T. G{\"o}rler, and F. Jenko, Phys. Rev. Lett. {\bf  106}, 055001 (2011).

\bibitem{SD} T. Stoltzfus-Dueck and B. Scott, Nucl. Fusion {\bf 57}, 086036 (2017).

\bibitem{Fowler-1964} T.K. Fowler, Phys. Fluids {\bf 7}, 249 (1964).

\bibitem{Fowler-1968} T.K. Fowler, {\em Thermodynamics of unstable plasmas}, in Advances in Plasma Physics
(ed. A. Simon and W. B. Thompson) , vol. 1, p. 201. New York: Interscience Press (1968).

\bibitem{Brizard-1991} A. Brizard, T.K. Fowler, D. Hua, and P.J. Morrison, Comm. Plasma Phys. Contr. Fusion {\bf 14}, 263 (1991). 

\bibitem{Biglari} H. Biglari, P.H. Diamond, and M.N. Rosenbluth, Phys. Fluids B {\bf 1}, 109 (1989).

\bibitem{Romanelli} F. Romanelli, Phys. Fluids B {\bf 1}, 1018 (1989).

\bibitem{Plunk} G.G. Plunk, P. Helander, P. Xanthopoulos, and J. W. Connor, Phys. Plasmas {\bf 21}, 032112 (2014).

\bibitem{Zocco} A. Zocco, P. Xanthopoulos, H. Doerk, J.W. Connor, and P. Helander, J. Plasma Phys. {\bf 84}, 715840101 (2018). 

\bibitem{Landreman-Plunk-Dorland} M. Landreman, G.G. Plunk and W. Dorland, J. Plasma Phys. {\bf 81}, 905810501 (2015). 

\end{references}

\end{document}